\documentclass[aip,jap,amsmath,amssymb,reprint,superscriptaddress,preprintnumbers,showpacs]{revtex4-1}
\pdfoutput=1
\usepackage[utf8]{inputenc}
\usepackage{bm,mathrsfs}
\usepackage[mathcal]{euscript}
\usepackage[breaklinks=true,unicode=true,urlcolor = blue,colorlinks = true,citecolor = blue,linkcolor = blue]{hyperref}
\usepackage{dcolumn}

\usepackage{graphicx} 				

\usepackage{color}
\usepackage{color}
\usepackage[normalem]{ulem}

\renewcommand{\vec}[1]{\bm{#1}}
\begin{document}


\title{Magnetization patterning induced by electrical spin-polarized current in nanostripes}

\author{Oleksii M. Volkov}
     \email{alexey@volkov.ca}
\affiliation{Taras Shevchenko National University of Kiev, 01601 Kiev, Ukraine}

\author{Volodymyr P. Kravchuk}
 \affiliation{Bogolyubov Institute for Theoretical Physics, 03680 Kiev, Ukraine}

\author{Denis D. Sheka}
\affiliation{Taras Shevchenko National University of Kiev, 01601 Kiev, Ukraine}

\author{Yuri~Gaididei}
 \affiliation{Bogolyubov Institute for Theoretical Physics, 03680 Kiev, Ukraine}

\author{Franz G. Mertens}
\affiliation{Physics Institute, University of Bayreuth, 95440 Bayreuth, Germany}

\date{\today}

%
%

\begin{abstract}
The combined action of a transverse spin-polarized current and the current-induced {\O}rsted field on long ferromagnetic nanostripes is studied numerically and analytically. The magnetization behavior is analyzed for stripes with various widths and for all range of the applied current density. It is established that {\O}rsted field does not destroy periodical magnetization structures induced by the spin-torque, e.g. vortex-antivortex crystal and cross-tie domain walls. However, the action of the {\O}rsted field disables the saturation state for the strong currents: a stationary state with a single longitudinal domain wall appears instead. Shape of this wall remains constant with the current increasing. The latter phenomenon is studied both numerically and analytically.

\end{abstract}


%
%

\maketitle

\section{Introduction}

A magnetic waveguide, which consists of periodic magnetic structures, in recent years becomes an object of interest due to Bragg reflection which affect the spin wave dispersion. Magnetic waveguides can be fabricated by alternating material~\cite{Wang09a,Ma11} or geometrical~\cite{Chumak09,Kim09a,Lee09,Huber13} parameters. All this magnetic waveguides are permanent, i.e. the spectra of spin waves cannot be changed dynamically after fabrication. However, it was demonstrated recently that using strong spin-polarized current one can induce periodical magnetization structures on demand in nanomagnets.~\cite{Volkov11,Gaididei12a,Kravchuk13,Volkov13e} These structures take the appropriate form according to the shape of the magnet and the current density: a square vortex--antivortex lattice is formed in a thin film,~\cite{Volkov11,Gaididei12a} a one--dimensional domain structure is formed in a nanowire~\cite{Kravchuk13} and intermediate vortex-antivortex structures are formed in a thin stripe.~\cite{Volkov13e} Such periodical magnetization structures induced by spin-polarized current can be used for dynamic control of spin wave spectra in low-power filters and in other magnonic devices.~\cite{Wang14a}

\begin{figure}
\includegraphics[width=0.9\columnwidth]{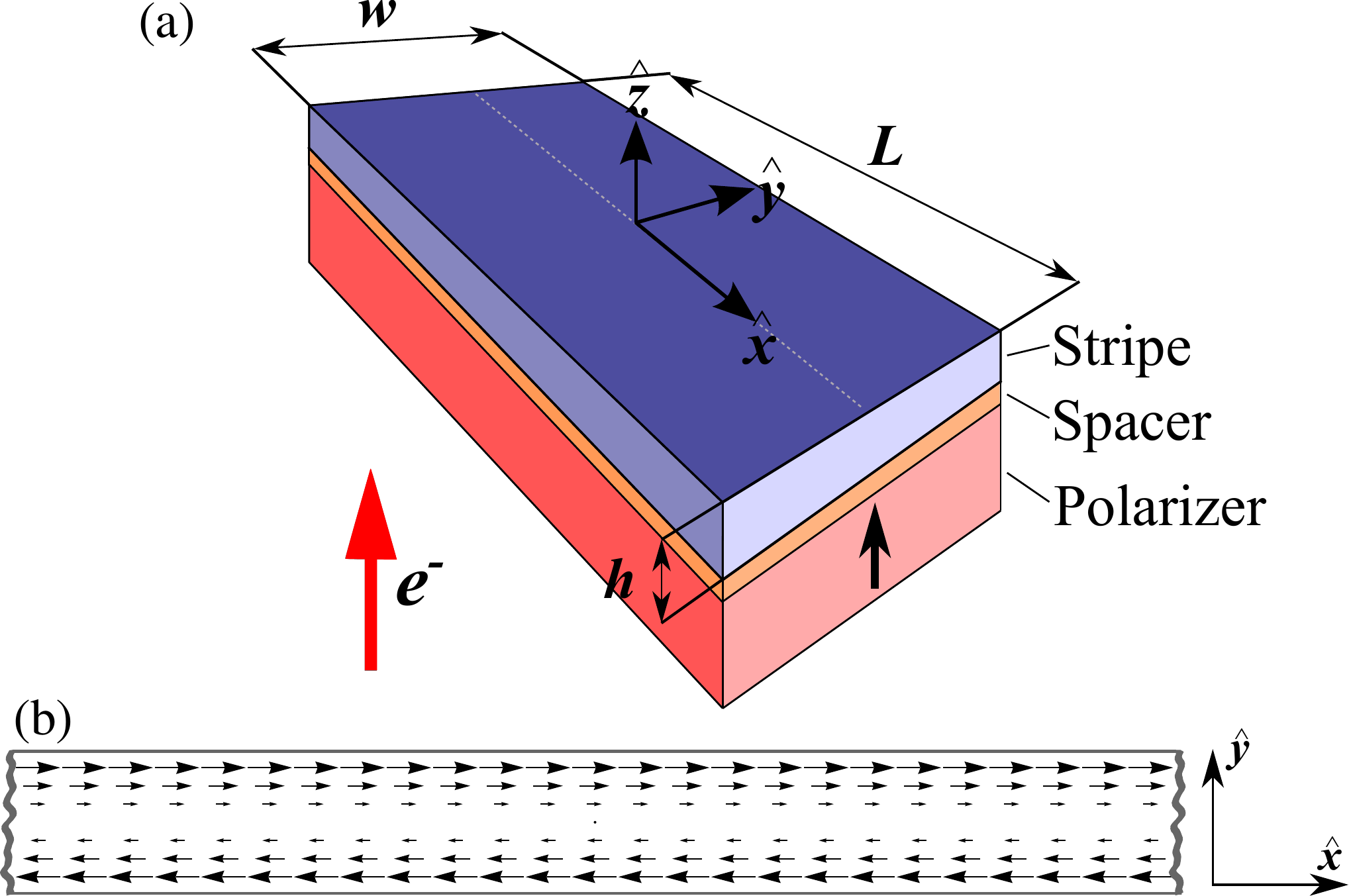}
\caption{(Color online) (a) The three--layer stripe--shaped spin valve. The spin polarized current flows perpendicularly to the studied stripe opposite to $\hat{z}$--direction, thereby the conduction electrons flow in the opposite side, as  shown by the red (large) arrow. The black (small) arrow indicates the direction of the polarizer magnetization. (b) The {\O}rsted field distributions induced by electrical current within the Sample plane for an infinite stripe length.}\label{fig:coords}
\end{figure}	

There are two main techniques which allow to inject pure spin-polarized current into a magnetic samples without creation of the current induced {\O}rsted field and significant heating. One of them utilizes geometrically separated areas with charge and spin currents injection and it is called non-local spin-current injection.~\cite{Jedema01,Jedema02,Ilgaz10,Jungwirth12,Motzko13} And another one utilizes spin-orbit torques which appear on the interconnection area of a ferromagnetic stripe and a nonmagnetic conductive layer with strong spin-orbit interactions. Using this spin-orbit torques one can create different realizations of classical field-like torques (e.g. indirect Rashba effect) and Slonczewski-Berger torque (e.g. spin Hall effects and the direct Rashba effect).~\cite{Takahashi08,Wei12,Khvalkovskiy13}

However, the simplest method of the spin-polarized current production is based on passing conducting electrons through a pillar magnetic heterostructure, see Fig.~\ref{fig:coords}~(a). The simplest pilar structure consists of two ferromagnetic layers (Polarizer and Sample) and nonmagnetic Spacer between them, see Fig.~\ref{fig:coords}~(a). When the electrical current passes through the Polarizer the conduction electrons become partially spin-polarized in a direction which is determined by the Polarizer magnetization. Polarizer is usually made of a hard ferromagnetic material whose magnetization is kept fixed. Spacer, being very thin (few nanometers), does not change spin polarization of the current electrons but it prevents the exchange and dipole-dipole interactions between Polarizer and Sample. Thus the spin-polarized electrons transfer the spin-torque from Polarizer to the Sample what can result in dynamics of the Sample magnetization.  The spin-torque influence can be described phenomenologically by  adding the SLT into Landau-Lifshitz equation.~\cite{Slonczewski96,Berger96,Slonczewski02} 

The electrical current generates {\O}rsted field, the exact form of which depends on a cross-section of the heterostructure. The aim of this work is to show that periodical structures can be formed in the case of combined action of a transverse spin-polarized current and the current-induced {\O}rsted field on long ferromagnetic nanostripes. In our study, we consider two different spatial distributions of the {\O}rsted field: for stripe with infinite and finite length, see Fig.~\ref{fig:coords}~(b) and Fig.~\ref{fig:Py_diagram_finite}~(a), respectively. By varying the stripe width we study the current induced magnetization behavior in wide range, starting from narrow stripes ($w\ll h$) and up to quasi two--dimensional wide stripes ($w\gg h$), where $w$ and $h$ denote the stripe width and thickness, respectively. We assume that the stripe is sufficiently long, so that $L\gg w$ and $L\gg h$ with $L$ being the stripe length, and thin enough to ensure uniformity of the magnetization along the thickness. Details of the problem geometry are shown in Fig.~\ref{fig:coords}.

\section{Model description}
	
Our study is based on the Landau--Lifshitz--Slonczewski phenomenological equation:~\cite{Slonczewski96,Berger96,Slonczewski02}
\begin{equation} \label{eq:LLS}
\dot{\vec{m}} = \vec m\times\dfrac{\delta\mathcal{E}}{\delta\vec{m}} - j \alpha \dfrac{\vec m\times[\vec m\times\hat{\vec{z}}]}{1+\beta (\vec{m}\cdot \vec{\hat{z}})},
\end{equation}
where $\vec{m}=\vec{M}/M_s=(m_x, m_y, m_z)$ is the normalized magnetization vector, $M_s$ is the saturation magnetization. The overdot indicates a derivative with respect to the rescaled time which is measured in units $(4 \pi \gamma M_s)^{-1}$, $\gamma$ is a gyromagnetic ratio and $\mathcal{E}=E/(4\pi M^2_s)$ is the normalized magnetic energy. The normalized spin--current density $j=J/J_0$, where $J_0=4 \pi M^2_s|e|h/\hslash$, with $e$ being the electron charge, $\hslash$ is the Planck constant. The spin--transfer torque efficiency coefficients $\alpha$ and $\beta$ have the forms $\alpha=P \Lambda^2/\left[ \Lambda^2 + 1\right]$ and $\beta=\left[ \Lambda^2 - 1\right]/\left[ \Lambda^2 + 1\right]$, where $P$ is the degree of spin polarization and the parameter $\Lambda$ describes the resistance mismatch between the spacer and the ferromagnet stripe.~\cite{Slonczewski02,Sluka11} The damping is omitted in Eq.~\eqref{eq:LLS}, because, as it was shown earlier,~\cite{Gaididei12a,Kravchuk13} the transverse spin--polarized current plays the role of an effective damping, which is usually greater than the natural one. It should also be noted that the Eq.~\eqref{eq:LLS} is written for the case when the Polarizer is magnetized along the $\hat{z}$--axis, see Fig.~\ref{fig:coords}.

We consider here a magnetic system, the total energy $E=E_\mathrm{ex}+E_\mathrm{d}+E_\mathrm{z}$ of which consists of three parts: exchange, dipole-dipole and Zeeman contributions. Exchange energy has the form
	\begin{equation} \label{eq:Eex}
		E_\mathrm{ex}=\frac{A}{2}\int_V \mathrm{d}\vec r~[(\nabla m_x)^2+(\nabla m_y)^2+(\nabla m_z)^2],
	\end{equation}
where $A$ is the exchange constant. 
	
The energy of dipole-dipole interaction is
	\begin{equation} \label{eq:Ems}
			E_\mathrm{d}=\frac{M_s^2}{2}\int_V\mathrm{d}\vec r\int_{V'}\mathrm{d}\vec{r}'~(\vec{ m}(\vec{r})\cdot\nabla)(\vec{m}(\vec r')\cdot\nabla')\frac{1}{|\vec r-\vec r'|}.
	\end{equation}
	
Zeeman energy describes the interaction of magnetic film with {\O}rsted field $\vec{B}(J,\vec{r})$ 
	\begin{equation} \label{eq:Ez}
		E_\mathrm{z}= - M_s \int_V \mathrm{d}\vec r ~\vec{B}(J,\vec{r}) \cdot \vec{m},
	\end{equation}		
where the spatial distribution of $\vec{B}(J,\vec{r})$ is determined by the form of the sample cross-section.

\section{Simulation results}
\label{sec:Simulation_results}

Here we report on the results of a numerical study which is based on the micromagnetic simulations.\footnote{We use the OOMMF code, version 1.2a5 for material parameters of Permalloy ($\mathrm{Ni}_{81}\mathrm{Fe}_{19}$): saturation magnetization $M_s=8.6 \times 10^5$~A/m, exchange constant $A=13 \times 10^{-12}$~J/m, and  anisotropy is neglected. Size of the mesh cell is $3 \times X \times h$~nm$^3$, where $X$ took values in interval from 0.5 to 3 nm, depending on the stripe width. The width is changed with steps $\Delta w=1$~nm for all samples. The current parameters are the following: polarization degree $P = 0.4$, and $\Lambda = 2$.}\cite{Donahue99}  The lengths of all studied stripes are the same $L=1\,\mathrm{\mu m}$. To ensure the magnetization uniformity along the $\hat{z}$--axis we consider only thin stripes with a thickness about one magnetic length, namely $h = 5$~nm. Since the thickness is small, the current density is assumed to be spatially uniform. The width is varied in a wide range $1\le w\le100$~nm.
A uniform in-plane magnetization state along the stripe (along the $\hat{x}$--axis) is chosen as an initial state for each simulation because it is very close to the ground state of a long stripe. In order to consider all possible current values we adiabatically increase the current density from zero to values where magnetization state does not depend on the current density. \footnote{Density of the applied current is changed accordingly to the law: $J=t~\Delta J/\Delta t$, where $\Delta J=10^{10}$~A/m$^2$ and $\Delta t=1$~ns. As a criterion of the saturation we use the relation $M_z/M_s>0.99$, where $M_z$ is the total magnetization along the $\hat{z}$--axis.}

\begin{figure*}
\begin{center}
\includegraphics[width=\linewidth]{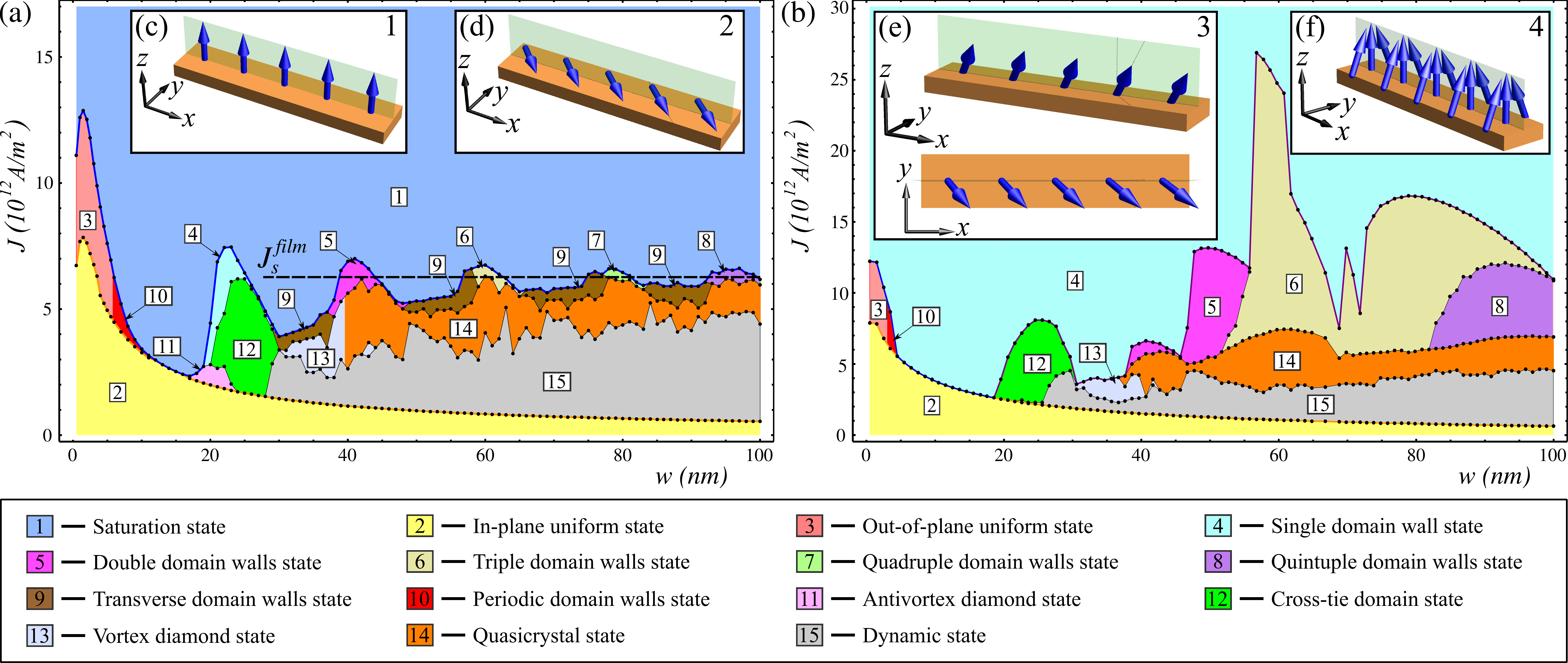}
\end{center}
\caption{(Color online) 
The phase diagrams of the magnetization behavior of Py stripes of different widths $w$ under the action of a transverse spin current $J$: (a) no {\O}rsted field, (b) {\O}rsted field is taken into account, see Fig~\ref{fig:coords}~(b), and periodic boundary conditions are applied.  Length $L=1$~$\mu$m and thickness $h=5$~nm of the stripes are fixed. Black bold dots indicate the transition from one state to another, and each state is numerated, in concordance with the legend. States 10-14 are illustrated in the Fig.~2 of the Ref.~\onlinecite{Volkov13e}. The inset pictures (c),(d) and (e) show the possible uniform states: the saturation state, the in-plane and the out-of-plane uniform states, respectively. The inset picture (f) shows the single domain wall state.
}\label{fig:Py_diagram}
\end{figure*}

Two different types of numerical experiments are performed by means of micromagnetic simulations. In the first type of simulations we consider finite length stripe samples under the action of pure spin-polarized current (without {\O}rsted field). As a result we obtain all magnetization states which were found in our previous studies for more thick samples.~\cite{Volkov13e} However, due to the thickness reducing from $h=10$~nm in the previous simulations to $h=5$~nm in the current ones, the typical current values, which corresponds to a certain magnetization state, become smaller. Namely, it is nearly three times less for the same stripe width, see Fig.~\ref{fig:Py_diagram}~(a) and Fig.~2 in Ref.~\onlinecite{Volkov13e}. Moreover, we perform the same simulations for stripes with periodic boundary conditions along the stripe which models the quasi-infinite stripe sample. In this simulations we do not find any principal differences with our previous results. This means that our stripe length $L=1$~$\mu m$ is large enough to generalize the phase diagram Fig.~\ref{fig:Py_diagram}~(a) for longer stripes.  

In the second type of simulations we consider quasi-infinite ferromagnetic stripe samples under the combined action of the spin-polarized current and the {\O}rsted field. In this simulations we also use the periodic boundary conditions along the stripe. In this case the exact form of the field reads: 
	\begin{equation} \label{eq:Oersted_field}
		\vec{B}(J,\vec{r})=\frac{4 \pi}{c} J y \vec{\hat{x}},
	\end{equation}
where $c$ is the speed of light. All possible types of the magnetization behavior in these micromagnetic simulations are summarized in the form of the phase diagram which is presented in Fig.~\ref{fig:Py_diagram}~(b). 

Comparing two diagrams, in Fig.~\ref{fig:Py_diagram}, one can conclude that the field influence is not significant for cases of narrow stripes and/or low current densities. This is due to the fact that the maximum value of {\O}rsted field is directly proportional to the current density and the stripe width $B_{max}\propto J w/2$. In these cases the same magnetization states appear: the uniform in-plane state for small current densities and for all stripe widths, see the region 2 of the phase diagrams in Fig.~\ref{fig:Py_diagram}; the uniform out-of-plane state for narrow widths and higher current densities, see the region 3; the periodic domain structure (region 10). However, for the case of strong currents the difference between these two cases is of principle: the action of the {\O}rsted field disables the saturation state and the single domain wall state appears instead of it, see the region 4 in Fig.~\ref{fig:Py_diagram}~(b) and the inset picture Fig.~\ref{fig:Py_diagram}~(f). This is the result of competition of influences of spin-torque and the {\O}rsted field on magnetization, and it will be discussed in more details in the Section~\ref{sec:Theory}. 

\begin{figure}
\begin{center}
\includegraphics[width=\linewidth]{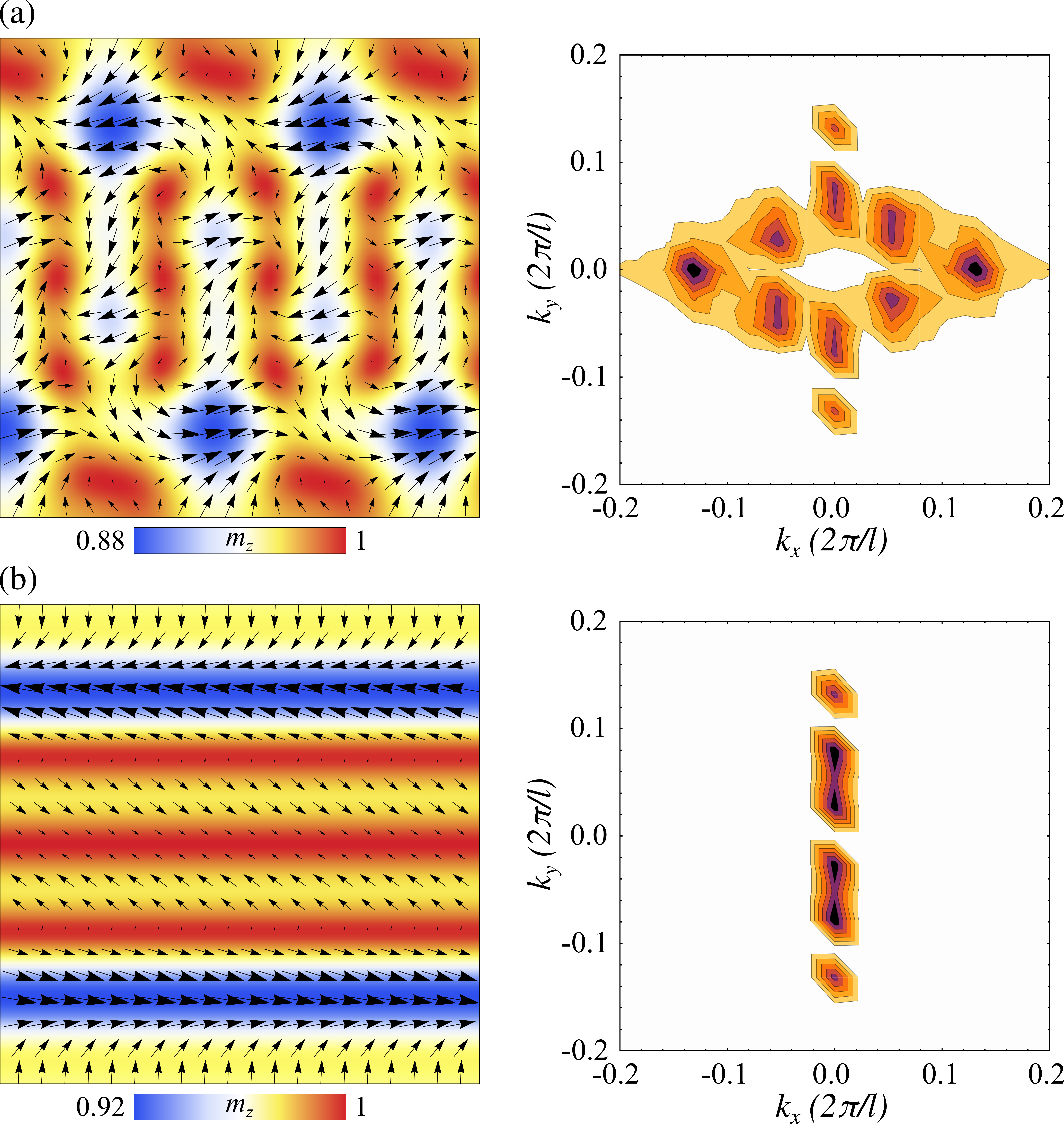}
\end{center}
\caption{(Color online) Magnetization distributions and its two-dimension Fourier spectrums of the square central part of the stripe with $w=93$~nm width with quasi-crystal magnetization state (a) and quintuple domain wall state (b), which were obtained for currents $J=6 \times 10^{12}$~A/m$^2$ and $J=7 \times 10^{12}$~A/m$^2$, respectively. The wave vectors $k_x$ and $k_y$ are measured in units $2\pi/\ell$, where $\ell$ is the exchange length. 
}\label{fig:2D_Fourier}
\end{figure}

For the intermediate values of the current densities the chaotic dynamical regime and stable periodical magnetization structures appear in both cases, with and without the {\O}rsted field. However, there is number of differences between these two regimes: (I) the vortex-antivortex  quasicrystals, the cross-tie domain walls and the vortex diamond states remain stable in both cases, although, in stripes under the action of the {\O}rsted field the quasicrystals undergo a deformation; (II) the antivortex diamond state and the transverse domain wall state do not remain under the {\O}rsted field action; (III) in the case of the field absence a state with single longitudinal domain wall appears in small region on the phase diagram between saturated and cross-tie domain wall states, see region 4 in Fig.~\ref{fig:Py_diagram}~(a). Whereas, in the case of joint action of the spin-current and the {\O}rsted field, the similar single domain wall state appears for any stripe width if the current is strong enough, see region 4 in Fig.~\ref{fig:Py_diagram}~(b); (IV) area of regions with multiple longitudinal domain walls increases significantly under the {\O}rsted field action.

\begin{figure}
\includegraphics[width=\columnwidth]{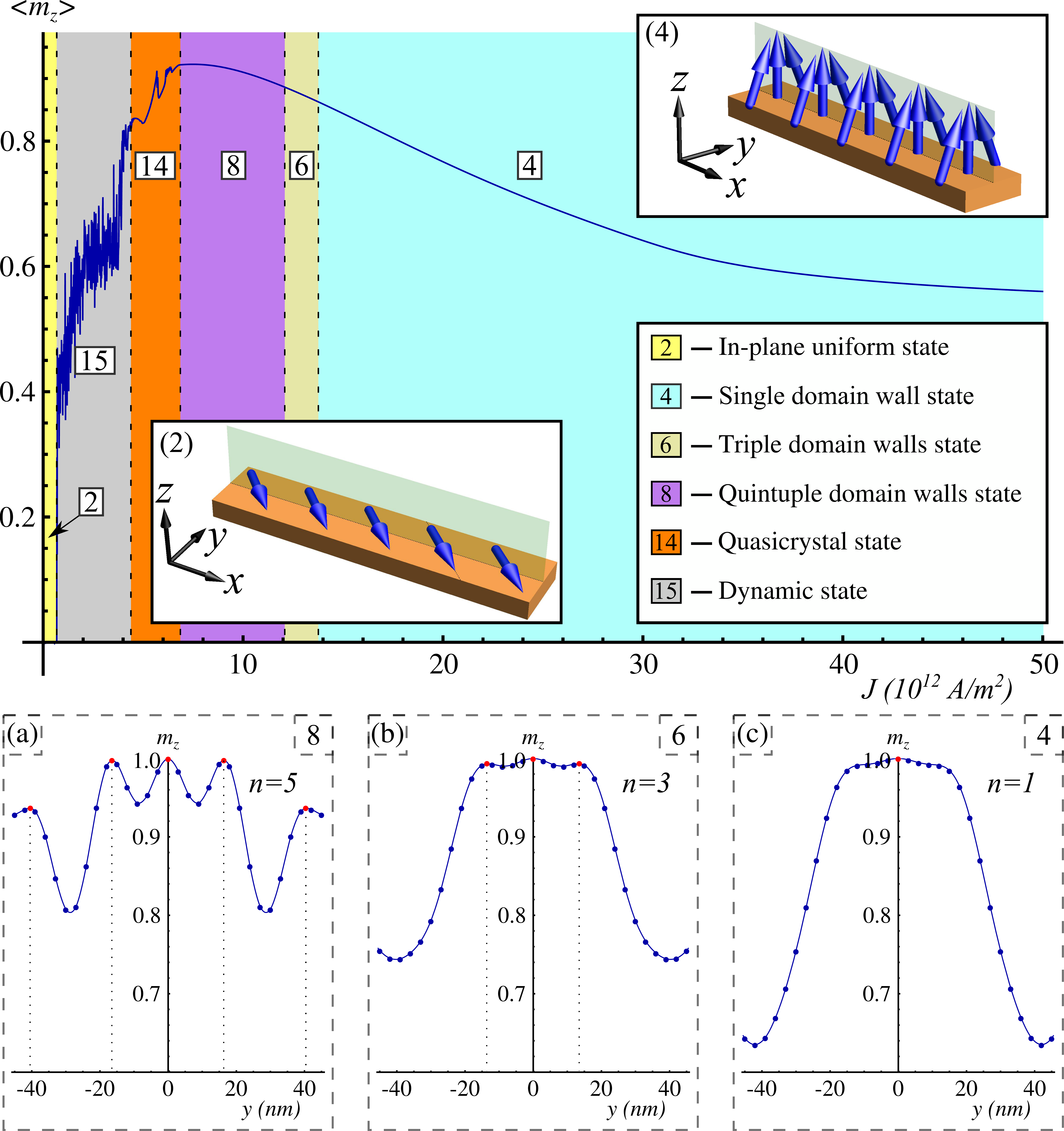}
\caption{(Color online) Dependence of the average out-of-plane magnetization component $\langle m_z \rangle$ on the current density in the case of the quasi-infinite stripe with $w=93$~nm. The row of insets (a), (b) and (c) show the out-of-plane magnetization distributions along the stripe width for quintuple ($J=7 \times 10^{12}$~A/m$^2$), triple ($J=12.5 \times 10^{12}$~A/m$^2$) and single ($J=15 \times 10^{12}$~A/m$^2$) longitudinal domain walls states, respectively. 
}\label{fig:1D_Fourier}
\end{figure}

We use two-dimensional (2D) Fourier transform of the out-of-plane magnetization component
\begin{equation} \label{eq:Fourier_2D}
	F^{2D}_z(\vec{k}_j)=\dfrac{1}{N_{xy}}\sum_{i=1}^{N_{xy}}\left[ m_z(\vec{r}_i)-\langle m_z \rangle \right] e^{-i \vec{k}_j\vec{r}_i }
\end{equation}  
to distinguish and analyze various periodical structures, such as longitudinal domain walls, quasicrystals and cross-tie domain walls. In Eq.~\eqref{eq:Fourier_2D},  $N_{xy}$ is the total number of mesh cells in the square area where Fourier transform is applied, $\vec{r}_i=(x_i,y_i)$ is a two-dimensional vector pointing to the appropriate cell of mesh, $\langle m_z \rangle=\frac{1}{N_{xy}}\sum_{i=1}^{N_{xy}} m_z(\vec{r}_i)$ is the averaged out-of-plane magnetization component and $\vec{k}_j=(k^x_j,k^y_j)$ is a 2D wave-vector. As one can conclude from the Fig.~\ref{fig:2D_Fourier}~(a), the action of the {\O}rsted field leads to the deformation of the quasicrystal state at the stripe edges, this results in formation of additional peaks in the 2D Fourier spectrum, see Fig.~\ref{fig:2D_Fourier}~(a). For the case of a longitudinal domain wall the $k_x$ components are absent in the Fourier spectrum, see Fih.~\ref{fig:2D_Fourier}~(b). This feature is used for the structure separation in the phase diagrams in Fig.~\ref{fig:Py_diagram}. On the other hand this feature allows us only to separate two-dimensional and one-dimensional magnetization structures but it couldn't help to separate the magnetization structures with different number of domain walls, which take place on both diagrams, see Fig.~\ref{fig:Py_diagram}~(a) and (b). For this separation we build the distribution of the out-of-plane magnetization component $m_z$ along the stripe width and we count all maximums which appear on it. This number of maximums shows to us the corresponding number of domain walls, as one can see form the inset pictures (a),(b) and (c) in Fig.~\ref{fig:1D_Fourier}.  

\begin{figure}
\begin{center}
	\includegraphics[width=1\columnwidth]{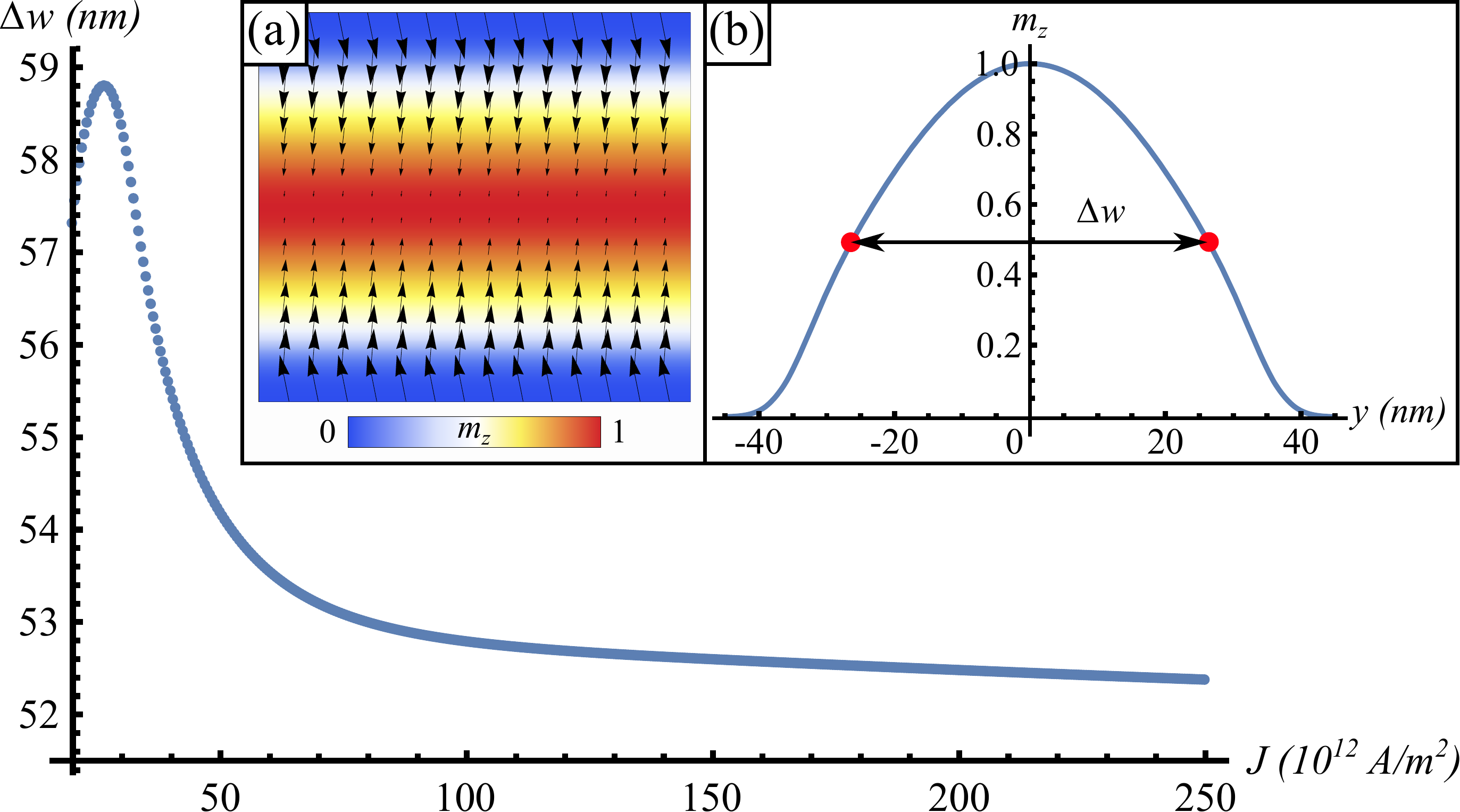}
\end{center}
\caption{(Color online) Dependences of the domain wall width $\Delta w$ on the current density for quasi-infinite Py stripe with $w=93$~nm for the case of combined action of the {\O}rsted field and spin-polarized current. 
The inset picture (a) shows the in-plane magnetization distribution in the central area of the stripe with dimensions $100\times 93$~nm$^2$ for the same current density. The inset picture (b) shows the out-of-plane magnetization distribution along the stripe width for $J=250\times10^{12}$~A/m$^2$ and definition of $\Delta w$.}\label{fig:FWHM}
\end{figure}

On the next stage of out studying we build the dependence of the averaged out-of-plane magnetization component $\langle m_z \rangle$ on the current density for stripe sample with $w=93$~nm, as one can see from the upper picture in Fig.~\ref{fig:1D_Fourier}. On this dependence one can see the regions of various magnetization states which appear during the current density increase: region 2 is the static homogeneous magnetization state within the plane of the stripe with $\langle m_z \rangle=0$ and it was described in details in Ref.~\onlinecite{Kravchuk13}; region 15 is the chaotic dynamical state of vortex-antivortex gas and it leads to a noisiness of the $\langle m_z \rangle$ component, however, its value grows with the current density; region 14 is the vortex-antivortex quasicrystal state, which was described in Ref.~\onlinecite{Gaididei12a}. The $\langle m_z \rangle(J)$ dependence is smooth but not monotonous. The existence of maximums is associated with structure reorganization of the quasicrystal, which occurs with the current growing; regions 8, 6 and 4 are magnetization states with five, three and one longitudinal domain walls, respectively. As one can see from the Fig.~\ref{fig:1D_Fourier} the $\langle m_z \rangle(J)$ dependence reaches its maximum in the region with five domain walls and after that it decreases smoothly with current increases. During this process the number of domain walls decreases to one. This happens due to the influence of the {\O}rsted field which becomes stronger with higher values of current density and it has the maximum values on the stripe edges, see Fig.~\ref{fig:coords}~(b). As a final magnetization state the single domain wall appears, it has its own characteristics: (i) in-plane magnetization components of the domain wall turn perpendicularly to the field, which is unusual, see the inset picture (a) in Fig.~\ref{fig:FWHM}; (ii) the profile of the domain wall is described by cosine in the center of stripe, see Fig.~\ref{fig:FWHM}~(b). It is different from the usual form of head-to-head or tail-to-tail domain walls in ferromagnetic samples which is described by hyperbolic secant; in addition (iii) the width of the domain wall $\Delta w$ is determined by stripe width and it does not depend on the material parameters as for usual domain walls. (iv) For large current densities the form of the domain wall becomes ``frozen'', in other words, it does not change during the current grows. To show the last characteristic we build the dependence of the domain wall $\Delta w$ on current density for stripe sample with $w=93$~nm width, see Fig.~\ref{fig:FWHM}. The width $\Delta w$ is found as a full width at half maximum for each current density, as it is shown on the inset of Fig.~\ref{fig:FWHM}~(b). As one can see, this dependence reaches the maximum and after that it decreases reaching some horizontal asymptote. This means that for infinitely high currents an unchangeable structure appears which is analyzed in Section~\ref{sec:Theory}.    

The maximum on the dependence on Fig.~\ref{fig:FWHM} can be explained as the influence of edge effects which appear from the competition of the dipole-dipole interaction with current induced field and spin-transfer torque on the stripe edge. The influence of the edge effects becomes even more stronger when we study the current action on finite length stripe samples, see Appendix~\ref{ap:Oersted-field-in-finite-stripe} for details.
       
%

\section{Longitudinal domain wall induced by strong current}
\label{sec:Theory}

In this section we show that for strong current densities the competition of the {\O}rsted field and the spin-torques results in a formation of the single longitudinal domain wall instead of the uniformly saturated state. For qualitative description of the phenomenon it is enough to model the magnetostatic energy of the stripe by the biaxial anisotropy:~\cite{Thiaville04,Thiaville05,Mougin07}
\begin{equation} \label{eq:Ean_eff}
	E^{\mathrm{ef}}_{\mathrm{an}} = \dfrac{1}{2} \int_V \mathrm{d}\vec r '~ \left(K_p m^2_z - K_a m^2_x\right),
\end{equation}   
where $K_p>0$ and $K_a>0$ are easy-plane and easy-axis anisotropy coefficients, respectively,
and can be assessed as demagnetization factors of thin ferromagnetic stripe.~\cite{Hubert98}   

Taking into account Eqs.\eqref{eq:Eex}, \eqref{eq:Ez},\eqref{eq:Oersted_field}, \eqref{eq:Ean_eff} and using the representation of the magnetization vector in the spherical coordinate system $\vec{m}=(\sin\theta \cos\phi; \sin\theta \sin\phi; \cos\theta)$ one can get the corresponding total normalized energy of the system:
\begin{equation} \label{eq:total_E}
	\begin{split}
	\mathcal{E}=\dfrac{1}{2}\int \mathrm{d}\vec{r} & \Big\{ \ell^2 \left[ (\nabla \theta)^2+ \sin^2 \theta (\nabla \phi)^2 \right]  + k_p\cos^2\theta \\ 
	& - k_a\sin^2\theta \cos^2\phi - 2 j \dfrac{y h}{s_0}\sin\theta\cos\phi \Big\},
	\end{split} 
\end{equation}
where  $k_p=\frac{K_p}{4 \pi M_s^2}$ and $k_a=\frac{K_a}{4 \pi M_s^2}$ are the normalized coefficients of effective anisotropy, $\ell=\sqrt{\frac{A}{4 \pi M_s^2}}$ is the exchange length and $s_0 = \frac{\Phi_0}{\pi B_0}$ is an effective area, where $\Phi_0=\hslash \pi c/|e|$ is magnetic flux quantum and $B_0=4 \pi M_s$ is the saturation field. For Permalloy $\ell\approx5.3$~nm, $B_0\approx1.08 \times 10^4$~G, it is remarkable that value of the effective area $s_0\approx 6.09\times 10^2$~nm$^2$ is of the same order of magnitude as area of the stripe cross-section.   
Substituting the energy \eqref{eq:total_E} into the equation \eqref{eq:LLS} and considering only static solution, one can write the set of equations:
\begin{subequations}
	\begin{align} 
		 \label{eq:system_1} \ell^2 \nabla \left( \sin^2\theta \nabla \phi \right) & -  k_a\sin^2\theta \sin\phi \cos\phi \nonumber \\
		 & -j \sin \theta \left[ \dfrac{yh}{s_0}\sin\phi + \dfrac{\alpha \sin\theta}{1+\beta \cos\theta} \right]=0, \\
		 \label{eq:system_2} \ell^2 \Delta \theta+ \dfrac{\sin 2\theta}{2} & \left[ k_p + k_a \cos^2\phi - \ell^2 (\nabla \phi)^2 \right] \nonumber \\
		 &+ j\dfrac{y h}{s_0}\cos\theta\cos\phi=0.
	\end{align}
\end{subequations}


One can see that in the case of high current density $k_p,k_a \ll jyh/s_0$ there is solution in linear approximation: 
\begin{subequations} \label{eq:linear}
	\begin{align} 
		\label{eq:lin_sol_theta} &\theta \approx \dfrac{h y}{s_0} \dfrac{2}{P} \left[1+O\left(\dfrac{h^2 w^2}{s^2_0}\right)\right], \\
		\label{eq:lin_sol_phi} &\phi \approx - \dfrac{\pi}{2}.
	\end{align}
\end{subequations}
The linear approximation in Eq.\eqref{eq:lin_sol_theta} works well in the whole range of parameters $y\in(0,w/2)$ under the following condition
\begin{equation} \label{eq:condition}
c = \frac{w^2h^2\left(1-\beta -2\beta ^2\right)}{24\alpha ^2s_0^2}\ll1.
\end{equation}
This means that one can neglect the next term in a series. For our material and geometrical parameters for stripe sample with $w=93$~nm one can estimate that $c\approx 0.08\ll1$, hence the linear dependence works well for all range of $y\in(0,w/2)$.
 
\begin{figure}
\begin{center}
	\includegraphics[width=1\columnwidth]{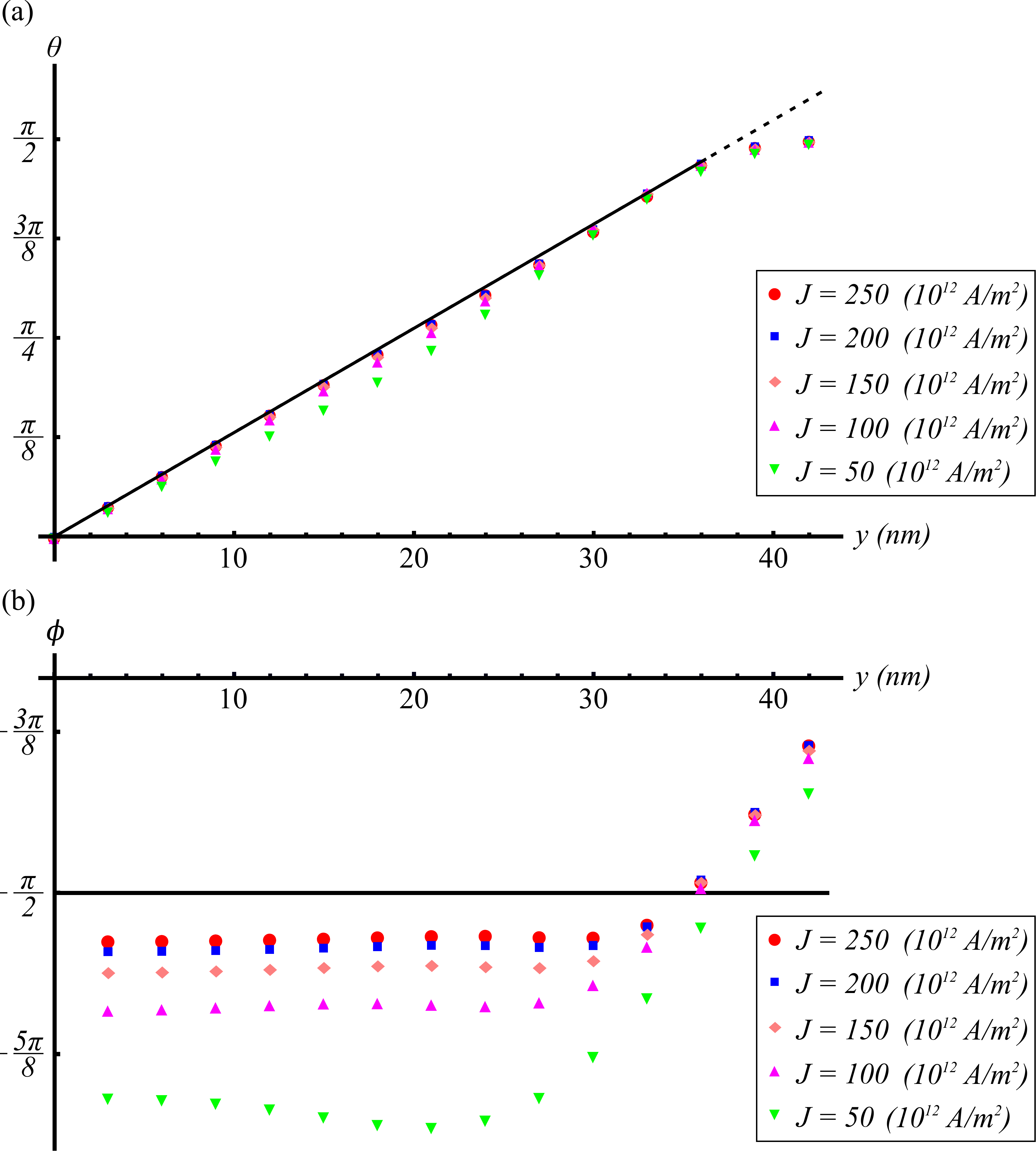}
\end{center}
\caption{(Color online) Analytical solutions \eqref{eq:lin_sol_theta} and \eqref{eq:lin_sol_phi} for the single domain wall state (solid lines) versus simulation data (point markers). Simulations were performed for Permalloy stripe sample with width $w=93$~nm. }\label{fig:Simulations_n_theory}
\end{figure}

The solution \eqref{eq:linear} originates from the competition of the spin-torque which is created by spin-polarized current and the influence of the {\O}rsted field. As one can see, solution \eqref{eq:lin_sol_theta} contains only geometrical and material parameters and it does not include current density. This fact means that for high current densities we obtain ``frozen'' single domain wall, whose form does not change with current increasing. As one can see from Fig.~\ref{fig:FWHM} and \ref{fig:Simulations_n_theory}, our analytical solutions \eqref{eq:linear} are in a good agreement with simulations data for large values of current density for strong enough currents
 
\section{Summary}
We study numerically the periodical structures formation under the combined action of spin-polarized current and the current-induced {\O}rsted field. In all studied cases, cross-tie, longitudinal domain walls and vortex-antivortex quasicrystals appear. As a result of competition of spin-polarized current and {\O}rsted field the single domain wall state in induced instead of the saturation, as in the case of pure spin current without the {\O}rsted field. It is shown both numerically and analytically that shape of this wall remains constant with the current increasing and it depends only on geometrical and material parameters of the sample. The micromagnetic simulations confirm our analytical results with high accuracy. 

\section*{Acknowledgements}

O.M.V. thanks the University of Bayreuth, where part of this work was performed, for kind hospitality. O.M.V acknowledges the support from DAAD funding program ``Research Grants for Doctoral Candidates and Young Academics Scientists'' (Code number 91530886-FSK).
	
\appendix

\section{The combined action of the spin-polarized current and the {\O}rsted field on stripe samples with finite length}
\label{ap:Oersted-field-in-finite-stripe}

\begin{figure}
\begin{center}
	\includegraphics[width=1\columnwidth]{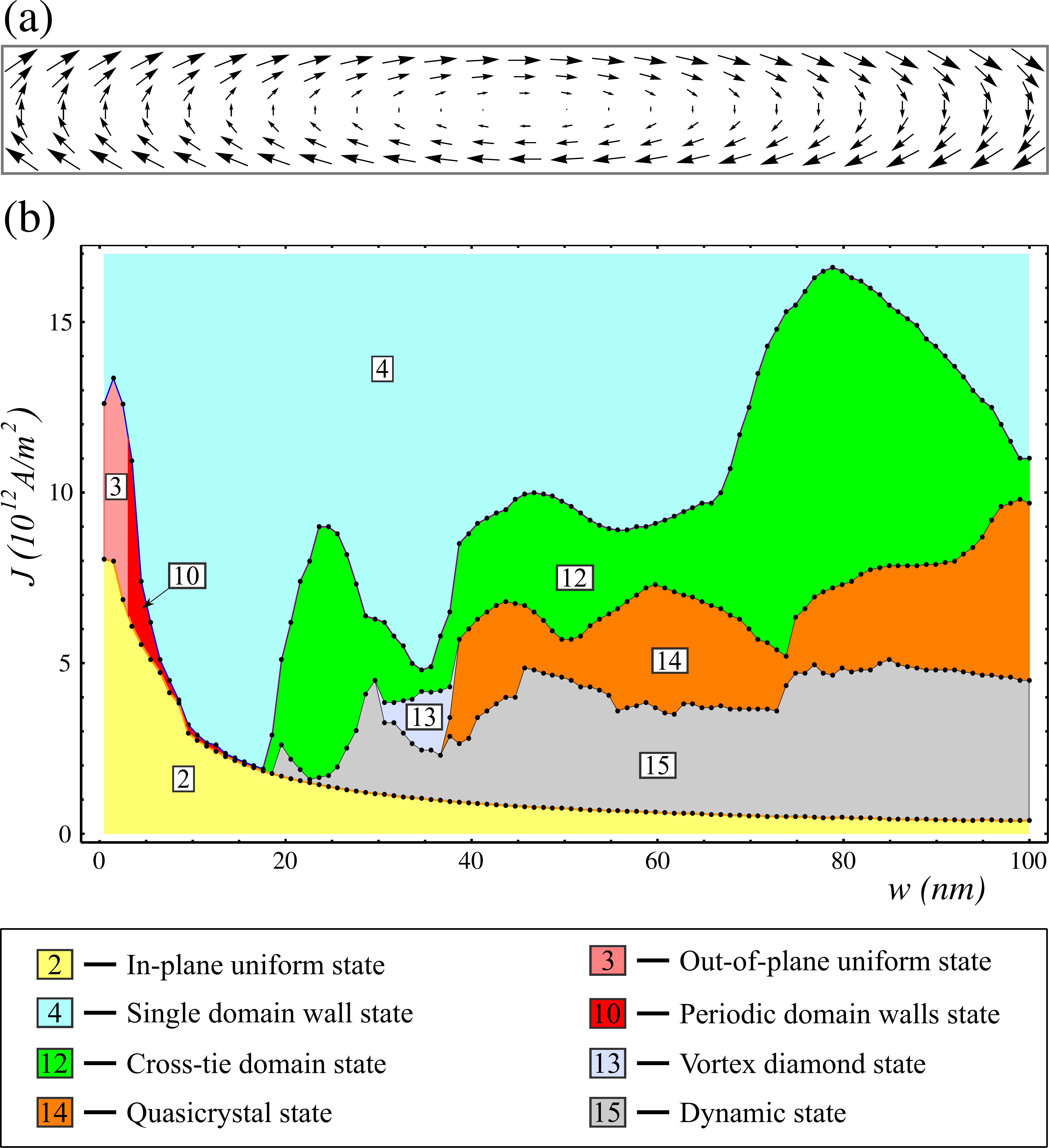}
\end{center}
\caption{(a) The spatial distribution of the {\O}rsted field induced by electrical current within the finite stripe sample. (b) The phase diagram of the magnetization behavior of Py finite length stripes of different widths $w$ under the co-action of the spin-polarized current and the {\O}rsted field, where all possible magnetization states have the same states numeration as they have had on Fig.~\ref{fig:Py_diagram}.}\label{fig:Py_diagram_finite}
\end{figure}

Here we report on the results of the co-action of the spin-polarized current and the {\O}rsted field on stripe samples with finite length. The spatial distribution of the {\O}rsted field for the finite length stripe can be calculated by using the Biot--Savart law:
\begin{equation}
	\vec{B}(J,\vec{r})=\frac{1}{c}\int_V \mathrm{d}\vec{r}' \frac{ \vec{J}\times (\vec{r}-\vec{r'})}{|\vec{r}-\vec{r'}|^3 }, 
\end{equation}
where $J$ is the current density and $|\vec{r}-\vec{r'}|=\sqrt{(x-x')^2+(y-y')^2+(z-z')^2}$ with $\vec{r}=(x,y,z)$ and $\vec{r'}=(x',y',z')$. Doing an integration over the entire volume with $x'\in(-\frac{L}{2},\frac{L}{2})$, $y'\in(-\frac{w}{2},\frac{w}{2})$ and $z'\in(-\infty,\infty)$ and taking into account that current flow throw the ferromagnetic stripe in $-\vec{\hat{z}}$-direction, while electrons are moving in the opposite direction, it gives to us the magnetic field which can be written in the complex vector $\vec{B}(r)=B_x+i B_y$ form with $\vec{\xi}=x+iy$:
\begin{equation} \label{eq:Finite_field}
	\vec{B}(J,\vec{\xi})= \dfrac{2 J}{c} \left\lbrace 2\pi y  + \sum^{4}_{k=1}(-1)^k \vec{\xi}^*_k \mathrm{ln}(\vec{\xi}^*_k) \right\rbrace
\end{equation}
where $\mathrm{ln}(\vec{\xi}_k)=\ln|\vec{\xi}_k|+i \arg(\vec{\xi}_k)$ is the complex logarithm, and  $\left\lbrace \vec{\xi}_1=(x+\frac{L}{2})+i(y-\frac{w}{2}) \right.$; $\vec{\xi}_2=(x-\frac{L}{2})+i(y-\frac{w}{2})$; $\vec{\xi}_3=(x-\frac{L}{2})+i(y+\frac{w}{2})$; $\left. \vec{\xi}_4=(x+\frac{L}{2})+i(y+\frac{w}{2}) \right\rbrace$. 

		
The final form of the {\O}rsted field distribution \eqref{eq:Finite_field} is described in Fig.~\ref{fig:Py_diagram_finite}~(a). As one can see, the central part of the field spatial distribution is the same to the {\O}rsted field in the infinite length stripe, however, the rest parts of them are completely different, as one can see from Fig.~\ref{fig:coords}~(b) and Fig.~\ref{fig:Py_diagram_finite}~(a), respectively.

In a result of simulations we obtain the phase diagram which is shown in Fig.~\ref{fig:Py_diagram_finite}~(b). As one can see, the parts of the diagram which correspond to the narrow stripes and small current densities remain almost the same as they appear in the previous cases which are discussed in section \ref{sec:Simulation_results} and, similarly to the case of quasi-infinite stripes under the action of the spin-current and {\O}rsted field, we also obtain the single domain wall state instead of saturated one. Meanwhile, the parts of the diagram in Fig.~\ref{fig:Py_diagram_finite}~(b) which correspond to the wide stripes and large current densities are completely different from the same parts of diagrams in Fig.~\ref{fig:Py_diagram}~(a) and (b). This occurs because the spatial distribution of the {\O}rsted field for finite length stripe leads to the strong influence of the edge effects in areas far from the center of the stripe. This edge effects start to play a key role in the processes of transition from one magnetization state to another: cross-tie domain wall appear instead of longitudinal domain walls with number of domains larger than one. At the same time, magnetization structures, e.g. cross-tie domain wall state, vortex diamond state and vortex-antivortex quasicrystals, remain stable, however, some of them undergo a deformation.

\end{document}